\documentclass[12pt]{article}

\usepackage{amssymb}
\usepackage{amsmath}
\numberwithin{equation}{section}

\newcommand{\prt}{\partial}

\newtheorem{propo}{Property}[section]

\newtheorem{defi}[propo]{Definition}
\newtheorem{theo}[propo]{Theorem}

\newcommand{\CC}{\mbox{${\mathbb C}$}}
\newcommand{\RR}{\mbox{${\mathbb R}$}}

\newcommand{\ZZ}{\mbox{${\mathbb Z}$}}

\newcommand{\II}{\mbox{${\mathbb I}$}}
\newcommand{\ca}{\mbox{$\mathcal{A}$}}
\newcommand{\cb}{\mbox{$\mathcal{B}$}}
\newcommand{\cc}{\mbox{$\mathcal{C}$}}
\newcommand{\cd}{\mbox{$\mathcal{D}$}}

\newcommand{{\cf}}{\mbox{$\mathcal{F}$}}
\newcommand{{\cg}}{\mbox{$\mathcal{G}$}}
\newcommand{\ch}{\mbox{$\mathcal{H}$}}

\newcommand{\ck}{\mbox{$\mathcal{K}$}}

\newcommand{\cR}{\mbox{$\mathcal{R}$}}

\newcommand{\cs}{\mbox{$\mathcal{S}$}}
\newcommand{\ct}{\mbox{$\mathcal{T}$}}
\newcommand{\cu}{\mbox{$\mathcal{U}$}}

\newcommand{\invdots}{{\mathinner{\mkern1mu\raise1pt
     \vbox{\kern7pt\hbox{.}}\mkern2mu\raise4pt\hbox{.}
     \mkern2mu\raise7pt\hbox{.}\mkern1mu}}}

\newcommand{\1}{\mathbf{1}}
\newcommand{\half}{\frac{1}{2}}

\begin{document}
\renewcommand{\thefootnote}{\fnsymbol{footnote}}

\pagestyle{empty}
\setcounter{page}{0}
\newcommand{\LAP}{LAPTH}
\def\logo{{\bf {\huge LAPTH}}}
\centerline{\logo}
\vspace {.3cm}
\centerline{{\bf{\it\Large 
Laboratoire d'Annecy-le-Vieux de Physique Th\'eorique}}}
\centerline{\rule{12cm}{.42mm}}
\vspace{20mm}
\begin{center}
  {\LARGE  {\sffamily Integrable systems with impurity}}
\\[2.1em]
{\large
E. Ragoucy\footnote{ragoucy@lapp.in2p3.fr}}
\\
\null
\noindent
\textit{LAPTH\footnote{UMR 5108 du CNRS associ\'ee \`a l'Universit\'e de 
Savoie},
9 Chemin de Bellevue, BP 110
F-74941 Annecy-le-Vieux  Cedex, France}
\end{center}
\vfill\vfill
\centerline{\textbf{Abstract:}}
    After reviewing some basic properties of RT algebras, which appear to be 
the natural framework to deal with integrable systems in presence of 
an impurity, we show how any integrable system (including these possessing 
translation invariance) can be promoted to an integrable system with 
an impurity which can reflect and transmit particles. The technics 
allows bulk translation invariant $S$-matrices while 
avoiding the no-go theorem stated recently about these laters.
\\[.21ex]
\begin{center}
Presented at the \textit{Vth International workshop on Lie theory and 
its applications in physics}, Varna 
(Bulgaria), June 16-22, 2003.
\end{center}
\vfill
\rightline{January 2004}
\rightline{LAPTH-Conf-1022/04}
\rightline{\tt math-ph/0401024}
\newpage
\pagestyle{plain}
\setcounter{footnote}{0}
\section{Introduction}
In the realm of factorized scattering theories in integrable models on the
line, the Zamolodchikov-Faddeev (ZF) algebra \cite{Zamolodchikov:xm, Faddeev:zy}
plays a fundamental role. It encodes (quadratically) the
 two-body scattering, which is the only particle interaction
present in this case. There have been various
generalizations, preserving integrability, to the case with a purely reflecting boundary 
\cite{Cherednik:vs,Sklyanin:yz,Ghoshal:tm,Liguori:de} or 
to the  case of point-like impurities
\cite{Delfino:1994nr, Castro-Alvaredo:2002fc}.
 In the later context, the relevant algebraic structure emerging recently
\cite{Mintchev:2002zd, Mintchev:2003ue,ZFrt}, is the so-called  
reflection-transmission (RT) algebra. It captures both
 the two-body scattering in the bulk, and the particle interaction 
with the impurity. 

After reviewing some basic properties of RT algebras, we 
present a procedure which allows, starting from any integrable model 
on the line, to built an integrable model with a point-like impurity. 
In particular, despite the property stated in 
\cite{Castro-Alvaredo:2002fc}, we show 
that the present technics produces theories possessing a (non trivial) translation 
invariant scattering matrix in the bulk together with an impurity which 
reflects and transmits particles.

\section{An example of reflection--transmission algebra}

To sustain the introduction of 
RT algebra in the study of quantum impurity problems, we start with the
simple example of (free) non-relativistic bosonic particles 
 on $\RR$, which interact with a $\delta$-type impurity 
localized in the origin.
This model is well-known to be exactly solvable (see e.g. \cite{A}). It has been 
presented from the RT algebra point of view in \cite{Mintchev:2003ue}, 
so that we just sketch the results, 
refering to the original paper for more details. In particular, we 
will consider here only the case $\eta\geq0$.

It is sufficient to investigate the spectral problem associated 
with  the one-particle Hamiltonian 
\begin{equation}
H^{(1)} = \sum_{i=1}^n - \frac{1}{2} \prt_{x_i}^2 + \eta\, \delta (x_i) \, ,
\qquad \eta \in \RR\, ,
\label{h1}
\end{equation}
 defined on a suitable domain $\cd_\eta$
of continuous functions on $\RR$, which are twice differentiable in
$\RR \setminus \{0\}$ and satisfy
\begin{equation}
\lim_{x\rightarrow 0^{+}}\left [(\prt_x \psi )(x) - (\prt_x \psi )(-x) \right ] =
2\eta\, \psi (0) \, .
\label{bc}
\end{equation}
$H^{(1)}$ is self-adjoint on $\cd_\eta$. A set $\Phi=\{\psi_k^\pm(x)\, ;\, k\in \RR \}$
of orthogonal (generalized) eigenstates, verifying (\ref{bc}), is 
provided by:
\begin{equation}
\psi_k^\pm(x) = \theta(\mp k) \left \{\theta(\mp x) T(\mp k) e^{ikx} +
\theta(\pm x)\left [e^{ikx} + R(\mp k)e^{-ikx}\right ] \right \}\, ,
\label{basis}
\end{equation}
where $\theta$ denotes the standard Heaviside function and 
\begin{equation}
T(k) = \frac{k}{k + i \eta}\, , \qquad
R(k) = \frac{-i\eta}{k + i \eta}\, .
\label{TR1}
\end{equation}
The family ${\overline \Phi}=\{{\overline \psi}_{-k}^{\, \pm}(x)\, ;\, k\in \RR \}$,
where the bar stands for complex conjugation, is also
orthonormal. The systems $\Phi$ and
${\overline \Phi}$ represent physically scattering
states and are separately complete in $L^2(\RR , dx)$.

The reflection and transmission from the impurity give rise to a
non-trivial scattering operator, which preserves the particle number. 
One can interpret $\psi_k^\pm (x)$
asymptotically as incoming waves, traveling in $\RR_\pm$
with momentum $k\not= 0$ towards the impurity. Accordingly, the vectors
\begin{equation}
|k\rangle^{\rm in} = \psi_k^+ (x) + \psi_k^- (x) \mbox{ and }
|k\rangle^{\rm out}  = {\overline \psi}_{-k}^{\, +} (x) +
{\overline \psi}_{-k}^{\, -} (x) 
\label{opin}
\end{equation}
form a basis of one-particle ``in"  states and ``out"  states resp.
The one-particle scattering operator is defined by
${\bf S}^{(1)}\,
|k\rangle^{\rm out} = |k\rangle^{\rm in}$.
By construction, ${\bf S}^{(1)}$ is a unitary operator on
$L^2(\RR , dx)$.  The one-particle transition amplitude reads
\begin{eqnarray}
{}^{\rm out}\langle p|k\rangle^{\rm in} =
{}^{\rm out}\langle p|{\bf S}^{(1)}|k\rangle^{\rm out} 
&=&[\theta (p) T(p) + \theta (-p) T(-p)]2\pi \delta (p-k) \nonumber \\
&&+[\theta (p) R(p) + \theta (-p) R(-p)]2\pi \delta (p+k) .\qquad
\label{opta}
\end{eqnarray}
Thus, $R$ and $T$ indeed represent the transmission and reflection amplitudes.
 
The $n$-particle amplitude, with initial and final configurations satisfying
$k_1<...<k_n$ and $p_1>...>p_n$ respect., can be expressed in
terms of (\ref{opta}):
\begin{equation}
{}^{\rm out}\langle p_1,...,p_n|k_1,...,k_n\rangle^{\rm in} =
\prod_{i=1}^n {}^{\rm out}\langle p_i|k_i\rangle^{\rm in} \, .
\label{ampl1}
\end{equation}
The basic feature of RT algebra is to provide an algebraic framework 
 for dealing with the above system when $\eta>0$,
similar to the familiar canonical commutation approach, 
which works in the case $\eta = 0$. 
 We introduce the associative
algebra $\cc$  with identity element $\bf 1$, generated by
$\{a^{\dag \xi} (k),\, a_\xi (k)\, :\, \xi=\pm,\, k\in \RR \}$
obeying the commutation relations:
\begin{eqnarray}
a_{\xi_1}(k_1)\, a_{\xi_2}(k_2) -  a_{\xi_2}(k_2)\, 
a_{\xi_1}(k_1) &=& 0\, ,
\label{ccr1} \\
a^{\dag \xi_1}(k_1)\, a^{\dag \xi_2}(k_2) - a^{\dag \xi_2}(k_2)\,
a^{\dag \xi_1}(k_1) &=& 0\, ,
\label{ccr2} \\
a_{\xi_1}(k_1)\, a^{\dag \xi_2}(k_2) - a^{\dag \xi_2}(k_2)\,
a_{\xi_1}(k_1) &=& 
\left [\delta_{\xi_1}^{\xi_2} + \ct_{\xi_1}^{\xi_2}(k_1)\right ] 2\pi
\delta(k_1-k_2)\, {\bf 1} \nonumber \\
&&+\cR_{\xi_1}^{\xi_2}(k_1) 2\pi \delta(k_1+k_2)\, {\bf 1}\, ,
\label{ccr3}
\end{eqnarray}
where
\begin{equation}
\ct(k) =\left(\begin{array}{cc} 0&T(k)\\ 
{\overline T}(k)&0\end{array}\right)\, ,
\qquad
\cR(k) =\left(\begin{array}{cc} R(k)&0\\0&
{\overline R}(k)\end{array}\right)\, .
\label{TR2}
\end{equation}
The right-hand side of eq. (\ref{ccr3}) captures the presence of the impurity.
The term proportional to $\delta(k_1+k_2)$ reflects in particular the
breaking of translation invariance due to the impurity.
We will see  that $\cc$ is a particular RT algebra. For
the moment we focus on the Fock representation $\cf$ of $\cc$, referring 
for the explicit construction to sect. \ref{fock}.
An essential feature of $\cf$ is that the operators 
$a^{\dag}(k)= (a^{\dag+} (k), a^{\dag-}(k))$ and 
$a(k)=\left(\begin{array}{c}a_+ (k) \\ a_- (k)\end{array}\right)$ 
in this representation satisfy
\begin{eqnarray}
a(k) &=& \ct(k) a(k) + \cR(k) a(-k) \, ,
\label{c1} \\
a^{\dag}(k) &=& a^{\dag}(k) \ct(k) +a^{\dag}(-k) \cR(-k) \, .
\label{c2}
\end{eqnarray}
The relations (\ref{c1}), (\ref{c2}) encode the interaction with the impurity.

The vacuum state $\Omega \in \cf$ obeys as usual $a_\xi (k)\Omega = 0$. We
denote by $( \cdot\, ,\, \cdot )$ the scalar product in $\cf$
and consider the vacuum expectation value
\begin{eqnarray}
&& (a^{\dag \epsilon_1}(p_1)...a^{\dag \epsilon_n}(p_n)\Omega
\, , \, a^{\dag \xi_1}(k_1)...a^{\dag \xi_n}(k_n)\Omega ) \, ,
\label{ampl2}\\
&& k_1<\cdots <k_n\, ,\quad \xi_i = -\epsilon (k_i)\, ,
\qquad p_1>\cdots >p_n\, ,\quad \epsilon_i = \epsilon (p_i) \qquad
\label{cond1}
\end{eqnarray}
$\epsilon$ being the sign function. By means of eqs. (\ref{ccr1})--(\ref{ccr3})
it is easily verified that (\ref{ampl2}) reproduces
the amplitudes (\ref{ampl1}) for any $n$.  Therefore, $\cc$ provides a purely
algebraic framework for constructing the scattering operator. 
{}In this framework, the Hamiltonian takes the form
\begin{equation}
H = \frac{1}{2} \int_{-\infty}^{+\infty} \frac{dk}{2\pi}k^2 a^{\dag
\xi}(k) a_\xi (k) \, .
\label{h2}
\end{equation}
The restriction of $H$ to the $n$-particle subspace 
 is the algebraic counterpart of the starting Hamiltonian.

Thus, $\cc$ is a universal and powerful tool for handling $\delta$-type impurities.
One can view $\cc$ as a central extension of the algebra of canonical
commutation relations. A direct generalization
is to substitute $\ct(k) {\bf 1}$ and $\cR(k) {\bf 1}$ in eq.
(\ref{ccr3}) with new generators
$t(k)$ and $r(k)$ which are no longer central. Moreover,
in the spirit of the ZF algebra, it is possible to
replace the bosonic exchange factor between $\{a^{\dag}(k),\, a(k)\}$ 
with a more general one. 

\section{The RT algebra}

Adopting throughout the paper the compact auxiliary spaces notation 
(see e.g. \cite{Ragoucy:2001zy}), we define RT algebras \cite{Mintchev:2003ue} by:
\begin{defi}{\bf (RT algebra $\cc_{S}$)}\\
A RT algebra is generated by $\1$ and the generators $A(k)$, 
$A^\dag(k)$, (which are column and line vector valued resp.) $t(k)$ and $r(k)$
(which are matrix valued) obeying:
\begin{eqnarray}
A_{1}(k_1)A_{2}(k_2) &=& S_{21}(k_{2},k_{1})
A_{2}(k_2)A_{1}(k_1)\label{rt-1}\\
A^\dag_{1}(k_1)A_{2}^\dag(k_2) &=&
A_{2}^\dag(k_2)A^\dag_{1}(k_1) S_{21}(k_{2},k_{1})
\label{rt-2}\\
A_{1}(k_1)A_{2}^\dag(k_2) &=&
A_{2}^\dag(k_2)S_{12}(k_{1},k_{2})A_{1}(k_1)+
\delta(k_1-k_2)\Big(\delta_{12} \1 +t_{12}(k_{1})\Big)
\nonumber\\
&& +r_{12}(k_1)\delta(k_1+k_2)\quad
\label{rt-3}\\
A_{1}(k_1)t_{2}(k_2) &=& S_{21}(k_{2},k_{1})
t_{2}(k_2)S_{12}(k_{1},k_{2}) A_{1}(k_1)
\label{rt-4}\\
A_{1}(k_1)r_{2}(k_2) &=& S_{21}(k_{2},k_{1})
r_{2}(k_2)S_{12}(k_{1},-k_{2}) A_{1}(k_1)
\label{rt-5}\\
t_{1}(k_1)A_{2}^\dag (k_2) &=&
A_{2}^\dag(k_2) S_{12}(k_{1},k_{2}) t_{1}(k_1)S_{21}(k_{2},k_{1})
\label{rt-6}\\
r_{1}(k_1)A_{2}^\dag (k_2) &=&
A_{2}^\dag(k_2) S_{12}(k_{1},k_{2}) r_{1}(k_1)S_{21}(k_{2},-k_{1})
\qquad
\label{rt-7}
\end{eqnarray}
\begin{eqnarray}
&& S_{12}(k_{1},k_{2}) t_{1}(k_1)S_{21}(k_{2},k_{1}) t_{2}(k_2)
\, =\, \qquad\nonumber\\ 
&& \qquad\qquad\qquad\qquad
 t_{2}(k_2) S_{12}(k_{1},k_{2}) t_{1}(k_1) S_{21}(k_{2},k_{1})\ 
\label{rt-8}\\
&& S_{12}(k_{1},k_{2}) t_{1}(k_1) S_{21}(k_{2},k_{1})
r_{2}(k_2) \, =\, \nonumber\\ 
&& \qquad\qquad\qquad\qquad
r_{2}(k_2) S_{12}(k_{1},-k_{2}) t_{1}(k_1) S_{21}(-k_{2},k_{1})
\label{rt-9}\\
&& S_{12}(k_{1},k_{2}) r_{1}(k_1) S_{21}(k_{2},-k_{1}) 
r_{2}(k_2) \, =\, \nonumber\\ 
&& \qquad\qquad\qquad\qquad
r_{2}(k_2) S_{12}(k_{1},-k_{2}) r_{1}(k_1) S_{21}(-k_{2},-k_{1})
\ \label{rt-10}
\end{eqnarray}
where $r(k)$ and $t(k)$ satisfy:
\begin{equation}
t(k)t(k)+r(k)r(-k) = \II \ \mbox{ and }\
 t(k)r(k)+r(k)t(-k) = 0 \label{rt-12}
\end{equation}
\end{defi}
We refer to $r(k)$ and $t(k)$ as reflection and transmission generators.
Here and below the $S$-matrix obeys the following well-known
Yang-Baxter equation and unitarity relation:
\begin{eqnarray}
&&S_{12}(k_{1},k_{2})\,S_{13}(k_{1},k_{3})\,S_{23}(k_{2},k_{3}) =
     S_{23}(k_{2},k_{3})\,S_{13}(k_{1},k_{3})\,S_{12}(k_{1},k_{2})
     \nonumber\\
&&S_{12}(k_{1},k_{2})\,S_{21}(k_{2},k_{1}) = \II\otimes\II
\label{uS}
\end{eqnarray}

A special case of RT algebra is the boundary algebra $\cb_{S}$ 
\cite{Liguori:1998xr} defined as the coset of $\cc_{S}$ by the relation $t(k)=0$. 
In the same way, $r(k)=0$ leads to an algebra with a purely transmitting
impurity \cite{Bowcock:2003dr}.

The generators $r(k)$ and $t(k)$ form a subalgebra $\ck_{S}\subset \cc_{S}$. 
This subalgebra contains itself two subalgebras:
 $\cR_{S}\subset \ck_{S}$, generated by $r(k)$, appears 
already in \cite{Sklyanin:yz}, called reflection algebra and 
$\ct_{S}\subset\ck_{S}$, generated by $t(k)$, called transmission algebra.

RT algebras possesses two important automorphisms: the (antilinear) adjunction $I$ 
given by
\begin{equation}
I \, : \,  \left\{\begin{array}{l}
a^{\dag}(\chi)  \mapsto  a(\chi)\ \mbox{ and }\
a(\chi)  \mapsto  a^{\dag}(\chi) \\
r(\chi) \mapsto r(-\chi)\ \mbox{ and }\ t(\chi) \mapsto t(\chi)
\end{array}\right.
\label{adj}
\end{equation}
and the reflection-transmission automorphism:
\begin{equation}
\varrho\, : \,  \left\{\begin{array}{l}
a^{\dag}(\chi)  \mapsto  a^{\dag}(\chi)\,t(\chi)+a^{\dag}(-\chi)\,r(\chi)\\
a(\chi)  \mapsto  t(\chi)\,a(\chi)+r(\chi)\,a(-\chi)\\
r(\chi) \mapsto r(\chi)\ \mbox{ and }\ t(\chi) \mapsto t(\chi)
\end{array}\right.
\label{rtinv}
\end{equation}
They both square to identity. We use them for the construction of 
Fock spaces for RT algebras.

\subsection{Fock spaces\label{fock}}
A Fock representation of the RT algebra with adjunction
$\{{\cc_S} , I\}$ is specified by the conditions:
\begin{itemize}
\item[{(i)}] the representation space is a complex Hilbert space $\ch$ with
scalar product $(\cdot\, ,\, \cdot )$;
\item[{(ii)}] the generators
$\{a(\chi ),\, a^{\dag} (\chi ),\, r(\chi ),\, t(\chi )\}$
are operator-valued distributions with common and invariant dense
domain $\cd\subset \ch$, where eqs. (\ref{rt-1})--(\ref{rt-12}) hold;
\item[{(iii)}] the adjunction $I$ is realized as a
conjugation with respect to $(\cdot\, ,\, \cdot )$;
\item[{(iv)}] there exists a vacuum state $\Omega \in \cd$,
which is annihilated by $a(\chi)$. The vector $\Omega $ is
cyclic with respect to $\{a^{\dag}(\chi)\}$ and
$(\Omega\, ,\, \Omega ) = 1$.
\end{itemize}

There is a number of simple, but quite important consequences from
the assumptions (i)--(iv). They are proved in \cite{Mintchev:2003ue}.
\begin{propo}\label{prop1} The reflection--transmission
automorphism $\varrho $
is realized in any Fock representation by the identity operator.
\end{propo}

Any RT algebra ${\cc_{S}}$ admits in general a whole
family $\cf(\cc_{S})$ of Fock representations, parameterized by
 the vacuum expectation values
\begin{equation}
\cR(\chi) = (\Omega\, ,\, r(\chi) \Omega )
\, , \qquad
\ct(\chi) = (\Omega\, ,\, t(\chi) \Omega ) \, ,
\label{vevTR}
\end{equation}
called transmission and reflection matrices. They obey:
\begin{propo} \label{prop2}
     In each Fock representation of $\{{\cc_{S}}, I\}$,
$\ct(\chi)$ and $\cR(\chi)$ satisfy the Hermitian
analyticity conditions
\begin{equation}
\cR^\dagger(\chi) = \cR(-\chi)\, ,\qquad
\ct^\dagger(\chi) = \ct(\chi)\, ,
\end{equation}
the consistency relations
\begin{eqnarray}
&&\cs_{12}(\chi_1 , \chi_2)\, \cR_1(\chi_1)\,
\cs_{21}(\chi_2 , -\chi_1)\, \cR_2(\chi_2) = \nonumber\\
&&\qquad \qquad \qquad \cR_2(\chi_2)\, \cs_{12}(\chi_1 , -\chi_2)\, \cR_1(\chi_1)\,
\cs_{21}(-\chi_2 , -\chi_1)\, ,
\label{rr1}\\
&&\cs_{12}(\chi_1 , \chi_2)\, \ct_1(\chi_1)\,
\cs_{21}(\chi_2 , \chi_1)\, \ct_2(\chi_2) = \nonumber\\
&&\qquad \qquad \qquad \ct_2(\chi_2)\, \cs_{12}(\chi_1 , \chi_2)\, \ct_1(\chi_1)\,
\cs_{21}(\chi_2 , \chi_1)\, ,
\label{tt1}\\
&&\cs_{12}(\chi_1 , \chi_2)\, \cR_1(\chi_1)\,
\cs_{21}(\chi_2 , -\chi_1)\, \ct_2(\chi_2) = \nonumber\\
&&\qquad \qquad \qquad \ct_2(\chi_2)\, \cs_{12}(\chi_1 , \chi_2)\,
\cR_1(\chi_1)\, \cs_{21} (\chi_2 , -\chi_1)
\label{tr1}
\end{eqnarray}
and unitarity conditions
\begin{eqnarray}
\ct(\chi) \ct(\chi) + \cR(\chi) \cR(-\chi) = \II\, ,\qquad
\ct(\chi) \cR(\chi) + \cR(\chi) \ct(-\chi) = 0 \, .\qquad
\label{unitTR0}
\end{eqnarray}
Moreover, the vacuum state $\Omega$ is unique (up to a phase
factor) and satisfies
\begin{equation}
r(\chi) \Omega = \cR(\chi) \Omega \, , \qquad \quad
t(\chi) \Omega = \ct(\chi) \Omega \, .
\label{e1}
\end{equation}
\end{propo}

We thus recover at the level of Fock representation the well-known
boundary Yang--Baxter equation (\ref{rr1}), together with a
 transmission (\ref{tt1}) and a transmission--reflection (\ref{tr1})
Yang--Baxter equation.

\section{Kinematics and RT algebras}
The goal of this section is to present a diagrammatic framework for
integrable models with impurity, which sustains the RT algebra 
presentation.
We parameterize the asymptotic particles by spectral parameter $k \in \RR$
and an ``isotopic" index $i=1,...,N$, corresponding to the auxiliary space.
We will use as fundamental building blocks  
 the two-body bulk scattering matrix 
$S_{12}(k_1,k_2)=
S_{i_1i_2}^{j_1j_2}(k_1,k_2)\,E_{i_{1}j_{1}}\otimes E_{i_{2}j_{2}}$;
the reflection matrix  $R(k )$, describing
the reflection of a particle from the impurity and
the transmission matrix $T(k )$, describing the
transmission of a particle by the impurity.

It is worth stressing that
$S$ is allowed to depend on $k_1$ and $k_2$ separately 
\cite{Liguori:de,Mintchev:2002zd,Mintchev:2003ue}.
Allowing $S$ to depend on
$k_1$ and $k_2$ separately, leads to a natural generalization 
of the inverse scattering method, which avoids the no-go theorem of
\cite{Castro-Alvaredo:2002fc} and
describes a large set of integrable systems, not covered there.
We will come back on this point in sect. \ref{impurity}.

It is convenient to define at this stage the matrices
\begin{equation}
R^\pm (k) \equiv \theta(\pm k) R(k) \, ,\qquad
T^\pm (k) \equiv \theta(\pm k) T(k) \, ,
\label{RT}
\end{equation}
which have a simple physical interpretation: $R^\pm$ describe the reflection
of a particle propagating $\RR_\pm$, whereas $T^\pm$ correspond to 
the transmission of a particle from $\RR_-$ to $\RR_+$ and vice versa, 
see fig. 1. Time is flowing along the vertical 
direction and single lines represent particles.
The double line corresponds to the impurity world line.

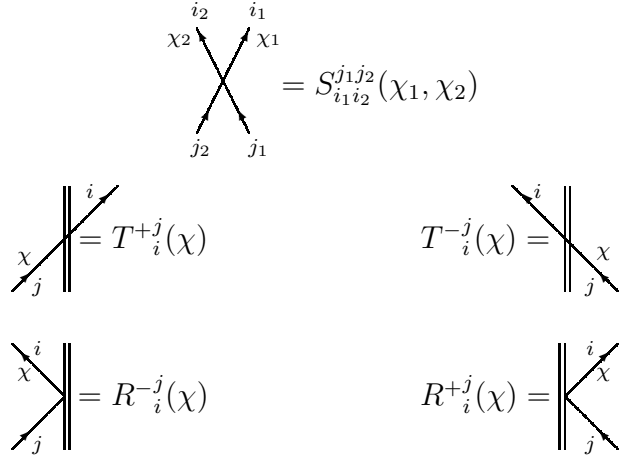
\begin{figure}[h]
\setlength{\unitlength}{0.7mm}
\begin{picture}(60,100)(-20,10)
\qbezier(65,80)(60,90)(55,100)
\qbezier(55,80)(60,90)(65,100)
\put(56.5,82.5){\vector(1,2){1}}
\put(64,97.5){\vector(1,2){1}}
\put(64,82.5){\vector(-1,2){1}}
\put(56.5,97.5){\vector(-1,2){1}}
\put(80,73){\makebox(20,20)[t]{$=S_{i_1i_2}^{j_1j_2}(\chi_1,\chi_2)$}}
\put(57,85){\makebox(20,20)[t]{${}_{i_1}$}}
\put(46,85){\makebox(20,20)[t]{${}_{i_2}$}}
\put(57,59){\makebox(20,20)[t]{${}_{j_1}$}}
\put(46,59){\makebox(20,20)[t]{${}_{j_2}$}}
\put(59,79){\makebox(20,20)[t]{${}_{\chi_1}$}}
\put(42,79){\makebox(20,20)[t]{${}_{\chi_2}$}}
\put (30,50){\line(0,0){20}}
\put (31,50){\line(0,0){20}}
\qbezier(20,50)(30,60)(40,70)
\put(22.5,52.5){\vector(1,1){1}}
\put(37.5,67.5){\vector(1,1){1}}
\put(35,44){\makebox(20,20)[t]{$={T^+}_i^j(\chi)$}}
\put(15,32.5){\makebox(20,20)[t]{${}_j$}}
\put(25,50.5){\makebox(20,20)[t]{${}_i$}}
\put(12.5,37.5){\makebox(20,20)[t]{${}_\chi$}}
\put (125,50){\line(0,0){20}}
\put (126,50){\line(0,0){20}}
\qbezier(135,50)(125,60)(115,70)
\put(132.5,52.5){\vector(-1,1){1}}
\put(118,67.5){\vector(-1,1){1}}
\put(100,44){\makebox(20,20)[t]{${T^-}_i^j(\chi)=$}}
\put(120,32.5){\makebox(20,20)[t]{${}_j$}}
\put(110,50.5){\makebox(20,20)[t]{${}_i$}}
\put(122.5,38.5){\makebox(20,20)[t]{${}_\chi$}}
\put (30,20){\line(0,0){20}}
\put (31,20){\line(0,0){20}}
\qbezier(20,20)(25,25)(30,30)
\qbezier(20,40)(25,35)(30,30)
\put(22.5,22.5){\vector(1,1){1}}
\put(22.5,37.5){\vector(-1,1){1}}
\put(35,14){\makebox(20,20)[t]{$={R^-}_i^j(\chi)$}}
\put(15,2.5){\makebox(20,20)[t]{${}_j$}}
\put(15,20.5){\makebox(20,20)[t]{${}_i$}}
\put(12.5,15.5){\makebox(20,20)[t]{${}_\chi$}}
\put (124,20){\line(0,0){20}}
\put (125,20){\line(0,0){20}}
\qbezier(135,20)(130,25)(125,30)
\qbezier(135,40)(130,35)(125,30)
\put(132.5,22.5){\vector(-1,1){1}}
\put(132.5,37.5){\vector(1,1){1}}
\put(100,14){\makebox(20,20)[t]{${R^+}_i^j(\chi)=$}}
\put(120,2.5){\makebox(20,20)[t]{${}_j$}}
\put(120,20.5){\makebox(20,20)[t]{${}_i$}}
\put(122.5,15.5){\makebox(20,20)[t]{${}_\chi$}}
\end{picture}
\caption{The two-body processes.}
\end{figure}

The construction of the possible three-body processes
in terms of $S$, $R^\pm$ and $T^\pm$ leads to a series of relations
\cite{Delfino:1994nr, Castro-Alvaredo:2002fc}.
Considering the scattering of three 
particles, we get the well-known quantum Yang-Baxter equation (in its braid form).
The consistency conditions implied by the scattering of two particles 
among each other and with the impurity are organized in the following 
three groups:

(a) pure reflection (fig. 2):
\begin{eqnarray}
&S_{12}(k_1, k_2)R^+_2(k_1)S_{12}(k_2 , 
-k_1)R^+_2(k_2) = \nonumber \\
&\qquad \qquad \qquad R^+_2(k_2)S_{12}(k_1, -k_2)R^+_2(k_1)S_{12}(-k_2, -k_1) \, ,
\label{SRSR+}\\
&S_{12}(k_1, k_2)R^-_1(k_2)S_{12}(-k_2 , 
k_1)R^-_1(k_1) = \nonumber \\
&\qquad \qquad \qquad R^-_1(k_1)S_{12}(-k_1, k_2)R^-_1(k_2)S_{12}(-k_2, -k_1) \, .
\label{SRSR-}
\end{eqnarray}
Eqs. (\ref{SRSR+}) and (\ref{SRSR-}) concern the reflection on $\RR_+$ and
$\RR_-$ respectively. Using the rules of fig. 1, one
gets from eq. (\ref{SRSR+}) the fig. 2. To 
any figure, one associates a new one, deduced  
 by reflection with respect to the impurity world line: eq. (\ref{SRSR-})
  gives the reflection of fig. 2.

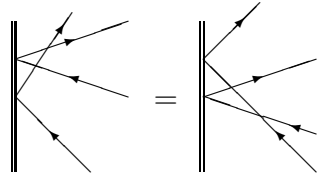
\begin{figure}[h]
\setlength{\unitlength}{.5mm}
\begin{picture}(50,40)(-60,0)
\put(35,0){\line(0,1){40}}
\put(34,0){\line(0,1){40}}
\put(65,20){\line(-3,1){30}}
\put(50,25){\vector(-3,1){1}}
\put(35,30){\line(3,1){30}}
\put(50,35){\vector(3,1){1}}
\put(55,0){\line(-1,1){20}}
\put(45,10){\vector(-1,1){1}}
\put(35,20){\line(2,3){15}}
\put(47,38){\vector(2,3){1}}
\put(65,5){\makebox(20,15)[t]{=}}
\put(85,0){\line(0,1){40}}
\put(84,0){\line(0,1){40}}
\put(115,10){\line(-3,1){30}}
\put(109,12){\vector(-3,1){1}}
\put(85,20){\line(3,1){30}}
\put(100,25){\vector(3,1){1}}
\put(115,0){\line(-1,1){30}}
\put(105,10){\vector(-1,1){1}}
\put(85,30){\line(1,1){15}}
\put(95,40){\vector(1,1){1}}
\end{picture}
\caption{Pure reflection.}
\end{figure}

(b) pure transmission (fig. 3):
\begin{eqnarray}
&&T^+_1(k_1)S_{12}(k_1, k_2)T^-_1(k_2) =
T^-_2(k_2)S_{12}(k_1, k_2)T^+_2(k_1) \, ,
\label{TST}\\
&&S_{12}(\chi_1, \chi_2)T^-_1(\chi_2)T^-_2(\chi_1) =
T^-_1(k_1)T^-_2(k_2)S_{12}(k_1, k_2)\, ,
\label{STT-}\\
&&S_{12}(\chi_1, \chi_2)T^+_1(\chi_2)T^+_2(\chi_1) =
T^+_1(k_1)T^+_2(k_2)S_{12}(k_1, k_2)\, ,
\label{STT+}
\end{eqnarray}

\begin{figure}[h]
\setlength{\unitlength}{.5mm}
\begin{picture}(40,40)(-25,-10)
\put(150,0){\line(0,1){40}}
\put(149,0){\line(0,1){40}}
\put(165,12){\line(-3,1){40}}
\put(156,15){\vector(-3,1){0.1}}
\put(160,0){\line(-1,1){30}}
\put(155,5){\vector(-1,1){0.1}}
\put(165,5){\makebox(20,15)[t]{=}}
\put(165,-10){\makebox(20,15)[b]{(b)}}
\put(200,0){\line(0,1){40}}
\put(199,0){\line(0,1){40}}
\put(225,10){\line(-3,1){35}}
\put(219,12){\vector(-3,1){0.1}}
\put(225,0){\line(-1,1){35}}
\put(220,5){\vector(-1,1){0.1}}
\put(25,0){\line(0,1){40}}
\put(24,0){\line(0,1){40}}
\put(5,0){\line(1,1){30}}
\put(10,5){\vector(1,1){0.1}}
\put(35,0){\line(-1,1){30}}
\put(30,5){\vector(-1,1){0.1}}
\put(40,5){\makebox(20,15)[t]{=}}
\put(40,-10){\makebox(20,15)[b]{(a)}}
\put(75,0){\line(0,1){40}}
\put(74,0){\line(0,1){40}}
\put(65,0){\line(1,1){30}}
\put(70,5){\vector(1,1){0.1}}
\put(95,0){\line(-1,1){30}}
\put(90,5){\vector(-1,1){0.1}}
\end{picture}
\caption{Pure transmission.}
\end{figure}
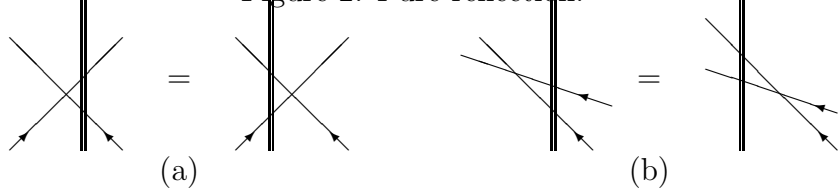

(c) mixed relations (fig. 4):
\begin{eqnarray}
&&R^+_1(k_1)T^-_2(k_2) =
T^-_2(k_2)S_{12}(k_1, k_2)R^+_2(k_1)S_{12}(k_2, -k_1) \, ,
\label{TSRS+}\\
&&T^+_1(k_1)R^-_2(k_2) =
T^+_1(k_1)S_{12}(k_1, k_2)R^-_1(k_2)S_{12}(-k_2, k_1) \, ,
\label{TSRS-}
\\
&&R^+_1(k_1)T^+_2(k_2) =
S_{12}(k_1, k_2)R^+_2(k_1)S_{12}(k_2, -k_1)T^+_2(k_2) \, ,
\label{SRST+}\\
&&T^-_1(k_1)R^-_2(k_2) =
S_{12}(k_1, k_2)R^-_1(k_2)S_{12}(-k_2, k_1)T^-_1(k_1)\, ,
\label{SRST-}
\end{eqnarray}
\begin{eqnarray}
&&R^+_1(k_1)T^-_2(k_2)S_{12}(-k_1, k_2) =
T^-_2(k_2)S_{12}(k_1, k_2)R^+_2(k_1) \, ,
\label{TSR+}\\
&&T^+_1(k_1)R^-_2(k_2)S_{12}(k_1, -k_2) =
T^+_1(k_1)S_{12}(k_1, k_2)R^-_1(k_2) \, ,
\label{TSR-}\\
&&R^+_2(k_1)S_{12}(k_2, -k_1)T^+_2(k_2) =
S_{12}(k_2, k_1)R^+_1(k_1)T^+_2(k_2) \, ,
\label{RST+}\\
&&R^-_1(k_2)S_{12}(-k_2, k_1)T^-_1(k_1) =
S_{12}(k_2, k_1)T^-_1(k_1)R^-_2(k_2) \, .
\label{RST-}
\end{eqnarray}

\begin{figure}[ht]
\setlength{\unitlength}{.5mm}
\begin{picture}(140,50)(10,-10)
\put(35,0){\line(0,1){40}}
\put(34,0){\line(0,1){40}}
\put(65,20){\line(-3,1){30}}
\put(50,25){\vector(-3,1){0.1}}
\put(35,30){\line(3,1){30}}
\put(50,35){\vector(3,1){0.1}}
\put(55,0){\line(-1,1){30}}
\put(50,5){\vector(-1,1){0.1}}
\put(65,5){\makebox(20,15)[t]{=}}
\put(65,-10){\makebox(20,15)[b]{(a)}}
\put(100,0){\line(0,1){40}}
\put(99,0){\line(0,1){40}}
\put(130,10){\line(-3,1){30}}
\put(121,13){\vector(-3,1){0.1}}
\put(100,20){\line(3,1){30}}
\put(115,25){\vector(3,1){0.1}}
\put(130,0){\line(-1,1){40}}
\put(125,5){\vector(-1,1){0.1}}
\end{picture}
%
\begin{picture}(120,50)(0,-10)
\put(35,0){\line(0,1){40}}
\put(34,0){\line(0,1){40}}
\put(50,0){\line(-1,2){15}}
\put(40,20){\vector(-1,2){0.1}}
\put(35,30){\line(1,2){5}}
\put(38.5,37){\vector(1,2){0.1}}
\put(55,0){\line(-1,1){30}}
\put(50,5){\vector(-1,1){0.1}}
\put(65,5){\makebox(20,15)[t]{=}}
\put(65,-10){\makebox(20,15)[b]{(b)}}
\put(100,0){\line(0,1){40}}
\put(99,0){\line(0,1){40}}
\put(110,0){\line(-1,2){10}}
\put(105,10){\vector(-1,2){0.1}}
\put(100,20){\line(1,2){10}}
\put(107,34){\vector(1,2){0.1}}
\put(120,10){\line(-1,1){30}}
\put(110,20){\vector(-1,1){0.1}}
\end{picture}
%
\begin{picture}(140,50)(10,-10)
\put(35,0){\line(0,1){40}}
\put(34,0){\line(0,1){40}}
\put(15,0){\line(1,1){40}}
\put(25,10){\vector(1,1){0.1}}
\put(65,0){\line(-3,1){30}}
\put(50,5){\vector(-3,1){0.1}}
\put(35,10){\line(3,1){30}}
\put(50,15){\vector(3,1){0.1}}
\put(65,5){\makebox(20,15)[t]{=}}
\put(65,-10){\makebox(20,15)[b]{(c)}}
\put(100,0){\line(0,1){40}}
\put(99,0){\line(0,1){40}}
\put(130,15){\line(-3,1){30}}
\put(118,19){\vector(-3,1){0.1}}
\put(100,25){\line(3,1){30}}
\put(112,29){\vector(3,1){0.1}}
\put(90,0){\line(1,1){40}}
\put(95,5){\vector(1,1){0.1}}
\end{picture}
%
\begin{picture}(120,50)(0,-10)
\put(35,0){\line(0,1){40}}
\put(34,0){\line(0,1){40}}
\put(47,1){\line(-1,2){12}}
\put(45,5){\vector(-1,2){0.1}}
\put(35,25){\line(1,2){8}}
\put(40,35){\vector(1,2){0.1}}
\put(20,0){\line(1,1){40}}
\put(30,10){\vector(1,1){0.1}}
\put(65,5){\makebox(20,15)[t]{=}}
\put(65,-10){\makebox(20,15)[b]{(d)}}
\put(100,0){\line(0,1){40}}
\put(99,0){\line(0,1){40}}
\put(80,0){\line(1,1){40}}
\put(90,10){\vector(1,1){0.1}}
\put(105,0){\line(-1,2){5}}
\put(103,4){\vector(-1,2){0.1}}
\put(100,10){\line(1,2){15}}
\put(114,38){\vector(1,2){0.1}}
\end{picture}
\caption{Mixed relations.}
\end{figure}
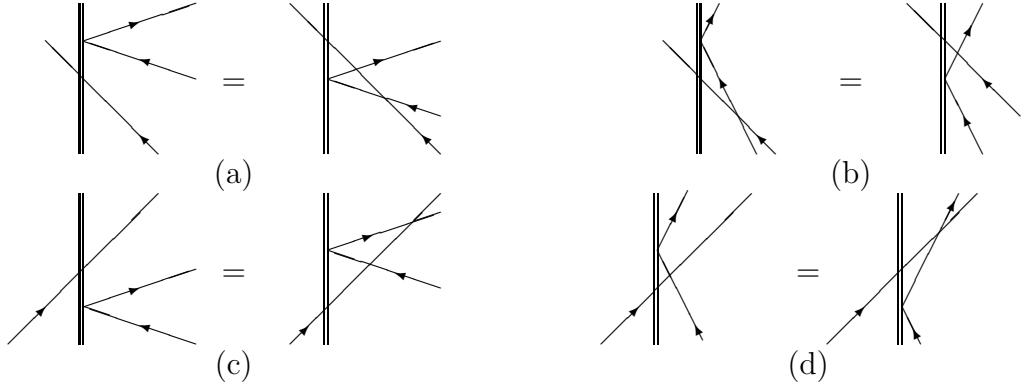

Finally, the requirements of unitarity and Hermitian analyticity lead for $S$ to
 the familiar \cite{Cherednik:vs, Sklyanin:yz, Ghoshal:tm} conditions
 (\ref{uS}).
Concerning $R$ and $T$, these conditions read
\begin{eqnarray}
T(k) T(k) + R (k) R (-k) = 1 \, ,&\qquad&
T (k) R (k) + R (k) T (-k) =0 \, ,\qquad 
\label{uTRTR}\\
{[T]}^\dagger (k) = T(k)\, ,&\qquad& {[R]}^\dagger (k) = R (-k)\, ,
\qquad 
\label{haTR}
\end{eqnarray}
The two sets of eqs. (\ref{TSRS+})-(\ref{SRST-}) and 
(\ref{TSR+})-(\ref{RST-}) being equivalent, 
one is left with the study eqs. (\ref{SRSR+}-\ref{SRST-}) and 
(\ref{uTRTR})-(\ref{haTR}).

The impurity at $x=0$ breaking translation invariance, we equip 
the generators of our algebra
with a double index $\alpha = (\xi , i)$, which amounts to extend the 
auxiliary space from $End(\CC^{N})$ to $End(\CC^{2N})$. The index $\xi = \pm$
indicates the half line $\RR_\pm$ where the particle is created or annihilated.
With this notation, we consider an associative algebra $\cc$ with 
identity element $\bf 1$, generated by $\{a(k ),\, a^{\dag}(k)\}$, 
which are subject to the constraints:
\begin{eqnarray}
&& a_{1}(k_1)\, a_{2 }(k_2) \, \; - \; \,
  \mathcal{S}_{21 }(k_2 , k_1)\, a_{2 }(k_2)\, a_{1 }(k_1) = 0
   \, , \qquad \label{aa}\\
&& a^{\dag}_1(k_1)\, a^{\dag}_2(k_2) - a^{\dag}_2(k_2)\, a^{\dag}_1(k_1)\,
  \mathcal{S}_{21 }(k_2 , k_1) = 0 \, , \qquad \label{a*a*} \\
&& a_{1 }(k_1)\, a^{\dag}_2(k_2) \; - \; a^{\dag}_2(k_2)\,
  \mathcal{S}_{12 }(k_1 , k_2)\, a_{1 }(k_1) =  \nonumber\\
&& \qquad\delta (k_1 - k_2) \left [\delta_{12 } + \ct_{12}(k_1)\right]\, {\bf 1} 
  + \delta (k_1 + k_2)\,  \cR_{12 }(k_1)\, {\bf 1}\, .
\label{aa*}
\end{eqnarray}
The constraints on the matrices $\cR(k)$ and $\ct(k)$ are given by 
(\ref{rt-9})-(\ref{rt-12}). When $\ct$ is invertible, they imply 
\cite{Mintchev:2002zd} the relations (\ref{SRSR+})-(\ref{SRST-}).

Then, the set $\{\mathcal{S}, \cR, \ct\}$ is related to the starting data 
$\{S,R,T\}$ of an RT algebra in the following way:
\begin{eqnarray}
\mathcal{S}_{\, (\xi_1,i_1)\, (\xi_2,i_2)}^{(\eta_1,j_1) (\eta_2,j_2)}
(k_1 , k_2) \equiv \delta_{\xi_1}^{\eta_2} \delta_{\xi_2}^{\eta_1} \,
S_{i_1 i_2}^{j_2 j_1 } (k_1 , k_2) \, ,
\label{calS}\\
\cR_{(\xi,i)}^{(\eta,j)}(k) \equiv \delta_\xi^\eta \, R_i^j(k) \, , \qquad
\ct_{\, \, (\xi,i)}^{(\eta,j)} (k ) \equiv \delta_{-\xi}^\eta \, T_i^j (k ) \, ,
\label{calRT}
\end{eqnarray}
where the auxiliary space indices have been explicited.

\section{From ZF to RT algebras}
We present here a connection between the ZF 
algebras and the RT algebras, in the same way boundary algebras 
(defined for integrable systems on the half-line 
\cite{Liguori:de,Liguori:1998xr}) can be 
constructed from ZF algebras \cite{Ragoucy:2001cf}. We remind 
that a ZF algebras $\ca_{S}$ is the polynomial algebra generated by a unit 
 $\1$ and the generators $a(k)$
and $a^\dag(k)$, subject to:
\begin{eqnarray}
a_{1}(k_1)a_{2}^\dag(k_2) &=&
a_{2}^\dag(k_2)S_{12}(k_{1},k_{2})a_{1}(k_1)+
\delta_{12}\delta(k_1-k_2) \1 \quad\\
a_{1}(k_1)a_{2}(k_2) &=& S_{21}(k_{2},k_{1})
a_{2}(k_2)a_{1}(k_1)\\
a^\dag_{1}(k_1)a_{2}^\dag(k_2) &=&
a_{2}^\dag(k_2)a^\dag_{1}(k_1) S_{21}(k_{2},k_{1})
\end{eqnarray}
We employ below an extension ${\overline\ca}_{S}$ of $\ca_{S}$,
involving power series in $a(k)$ and $a^\dag(k)$ whose individual terms
preserve the particle number. The concept of extended ZF (EZF) algebra
${\overline\ca}_{S}$ is relevant for proving \cite{Ragoucy:2001zy}:
\begin{propo}{\bf (Well-bred operators)}\\
Each ${\overline\ca}_{S}$ contains an invertible element $L(k)$ satisfying
\begin{eqnarray}
L_{1}(k_{1})a_{2}(k_{2})&=&
S_{21}(k_{2},k_{1})a_{2}(k_{2})L_{1}(k_{1}) \\
L_{1}(k_{1})a^\dag_{2}(k_{2}) &=&
a^\dag_{2}(k_{2})S_{12}(k_{1},k_{2})L_{1}(k_{1})\\
S_{12}(k_{1},k_{2})L_{1}(k_{1})L_{2}(k_{2}) &=&
L_{2}(k_{2})L_{1}(k_{1})S_{12}(k_{1},k_{2})
\end{eqnarray}
Hence, $L(k)$ generates a quantum group $\cu_{S}\subset{\overline\ca}_{S}$.
\end{propo}
$L(k)$, called well-bred operator, is explicitly constructed in
\cite{Ragoucy:2001zy} and admits a series representation in terms of 
$a(k)$ and $a^\dag(k)$.

We now turn to the construction of RT algebras from ZF ones.
Let $R(k)$ and $T(k)$ be any two matrix functions satisfying
\begin{equation}
R^\dagger(k)\, =\, R(-k)\, , \qquad T^\dagger(k) = T(k)\,,
\end{equation}
the unitary conditions (\ref{rt-12}) and the boundary Yang-Baxter equation (\ref{rt-10}).
It has been shown in \cite{Mintchev:2003ue} that $R(k)$ and $T(k)$ 
obey automatically the transmission and reflection-transmission Yang-Baxter eqs.
(\ref{rt-8})-(\ref{rt-9}).
Then, starting from the EZF algebra ${\overline\ca}_{S}$, one proves 
\cite{ZFrt}:
\begin{propo}
Let
$t(k) = L(k)T(k)L^{-1}(k)$ and $r(k) = L(k)R(k)L^{-1}(-k)$,
where $L(k)$ is the well-bred operator of ${\overline\ca}_{S}$ and 
$R(k)$ and $T(k)$
are defined as above. Then, $a(k)$, $a^\dag(k)$, $t(k)$ and
$r(k)$ obey the relations {\rm (\ref{rt-4})--(\ref{rt-12})}.
\end{propo}
\begin{propo}
The map
\begin{equation}
\sigma\left\{\begin{array}{lll}
a(k) & \rightarrow & \alpha(k)\, =\, t(k)a(k)+r(k)a(-k)\\
a^\dag(k) & \rightarrow &\alpha^\dag(k)\, =\, a^\dag(k)t(k)+a^\dag(-k)r(-k)
\end{array}\right.
\end{equation}
extends to a homomorphism on ${\overline\ca}_{S}$.
\end{propo}

The homomorphism $\sigma$ is essential for constructing an RT algebra from
EZF. Indeed, we have:
\begin{theo}\label{theo1}
Let
\begin{eqnarray}
A(k) &=& \half\Big(a(k)+\alpha(k)\Big) =\half\Big(
[1+t(k)]a(k)+r(k)a(-k)\Big)\\
A^\dag(k) &=& \half\Big(a^\dag(k)+\alpha^\dag(k)\Big)=\half\Big(
a^\dag(k)[1+t(k)]+a^\dag(-k)r(-k)\Big)\qquad 
\end{eqnarray}
Then $A(k)$, $A^\dag(k)$, $t(k)$ and $r(k)$ form a RT algebra.

In this construction,
$\ck_{S}$ becomes a Hopf coideal in $\cu_{S}$.
\end{theo}
Note that if one imposes on $R(k)$ the additional relation
$R(k)R(-k)=\II$, one recovers the construction
\cite{Ragoucy:2001cf} of the boundary algebra $\cb_S$. 

\subsection{Hamiltonians}

One can associate to any ZF algebra a hierarchy of Hamiltonians
\begin{equation}
H^{(n)}_{ZF}=\int_{\RR}dk\, k^n\, a^\dag(k)a(k),\ n\in\ZZ_{+}
\end{equation}
They form an Abelian set and
admit as symmetry algebra the quantum group generated by the
well-bred operators \cite{Ragoucy:2001zy}:
\begin{equation}
[H^{(n)}_{ZF},L(k)]=0,\ \forall n
\end{equation}

In a similar way, for any RT algebra, one introduces \cite{ZFrt}:
\begin{equation}
H_{RT}^{(n)}=\int_{\RR}dk\, k^n\, A^\dag(k)A(k)
\end{equation}
which obey to the following relations:
\begin{equation}
[H^{(m)}_{RT},H^{(n)}_{RT}]= [(-1)^m - (-1)^n] \int_{\RR}dk\, k^{m+n}\, \,
A^\dag(k)r(k)A(-k) , \label{comHRT}
\end{equation}
Thus, the Hamiltonians of the same parity define a hierarchy corresponding to 
integrable systems with impurity.
\begin{propo}
The subalgebra $\ck_{S}$ is a symmetry algebra of the hierarchy
$H_{RT}^{(n)}$:
$[H^{(n)}_{RT},t(k)]=0$ and $[H^{(n)}_{RT},r(k)]=0$.
\end{propo}
This result provides an universal model-independent description of
the symmetry content of the hierarchy $H_{RT}^{(n)}$. 
Remark that $r(k)$ and $t(k)$ encode both the particle-impurity
interactions and the quantum integrals of motion. This property of 
$\ck_{S}$ has been
applied in \cite{Mintchev:2001aq} for studying the symmetries of
the nonlinear Schr\"odinger equation on the half line.
 
Let us observe in conclusion that there is a simple relation \cite{ZFrt} between 
the hierarchies with and without impurity:
\begin{equation}
H_{RT}^{(n)}=H_{ZF}^{(n)}+ \int_{\RR}dk\,k^n\,
a^\dag(k)\Big(r(k)a(-k)+t(k)a(k)\Big)
\label{relation}
\end{equation}
It extends the 
results given in \cite{ZFconf} in the case of purely reflecting boundary.

\newpage
\section{Including an impurity on the line\label{impurity}}
It is natural to expect the existence of an integrable system with 
impurity associated to (almost) any integrable system defined on the 
whole line. However, the class of integrable systems 
possessing translation invariant $S$-matrix seems to be excluded
\cite{Castro-Alvaredo:2002fc}. The aim of this 
section is to show how RT algebras can circumvent this no-go theorem. 
Starting from 
an integrable system defined on a line and possessing a translation 
invariant $S$-matrix, we present here a general construction that allows
to obtain an integrable system, associated to the original one, 
 but with an impurity located at the origin which reflects 
and transmits particles. 

Let $s_{12}(k_{1}-k_{2})\in End(\CC^{N})\otimes End(\CC^{N})$ be the 
translation invariant $S$-matrix of the initial system. 
From this $S$-matrix, we define a new $S$-matrix:
\begin{equation}
 \!\!\!    
 S(k_{1},k_{2})=\left(
    \begin{array}{cccc}
  s_{12}(k_{1}-k_{2}) & 0 & 0 & 0 \\
  0 & s_{12}(k_{1}+k_{2}) & 0 & 0 \\
  0 & 0 & s_{12}(-k_{1}-k_{2}) & 0 \\
  0 & 0 & 0 & s_{12}(k_{2}-k_{1}) 
   \end{array}\right)
   \label{S-impure}
\end{equation}
which belongs to $End(\CC^{2N})\otimes End(\CC^{2N})$.
It is a simple matter of calculation to show that this $S$-matrix 
also obey the Yang-Baxter equation and unitary condition (\ref{uS}). 
To this $S$-matrix we associate a 
RT-algebra with generators $t(k)$, $r(k)$ and
\begin{equation}
    A(k)=\left(\begin{array}{c} a_{+}(k) \\ a_{-}(k)   \end{array}\right)
\ \mbox{ , }\ 
    A^\dag(k)=\left( a^\dag_{+}(k), a^\dag_{-}(k)\right)
\end{equation}
where $a_{\pm}(k)$ and $a^\dag_{\pm}(k)$ are $N$-vectors (column and 
line respect.), so that $A_{\pm}(k)$ and $A^\dag_{\pm}(k)$ are 
$2N$-vectors. The defining relations for the RT-algebra are those 
given in (\ref{rt-1})-(\ref{rt-10}), with the $S$-matrix (\ref{S-impure}). 

The physical interpretation of such a system is clear: the generators 
$a_{\pm}(k)$ and $a^\dag_{\pm}(k)$ correspond to annihilation and
creation operators in $\RR_{\pm}$.  They interact among themselves 
with the translation invariant $s_{12}(k_{1}-k_{2})$ matrix, in 
accordance with the naive picture that "far from the impurity", the 
translation invariance should not be broken. However, the 
exchange of "+"  with "-" operators (which must cross the impurity)
is done with the matrix 
$s_{12}(k_{1}+k_{2})$ which is \underline{not} translation invariant. 
Consequently, the complete matrix $S(k_{1},k_{2})$ is not 
translation invariant either, so that the property stated in \cite{Castro-Alvaredo:2002fc} 
does not apply here, as it was pointed out in \cite{Mintchev:2002zd}. Note also 
that despite the translation invariance, the generators 
$\{a_{\xi}(k),\,a^\dag_{\xi}(k)\}$ ($\xi=+$ or $-$)
does not obey ZF algebra relations, but rather 
boundary algebra ones. This is not surprising, this latter algebra 
being introduced for systems defined on the half-line \cite{Liguori:de}.

In view of the construction, one can look for generators of the form
\begin{equation}
    t(k)=\left(\begin{array}{cc}
     0 & \tau(k) \\ \tau(-k) & 0 
   \end{array}\right)
   \ \mbox{ and }\ 
    r(k)=\left(\begin{array}{cc}
  \rho(k) & 0 \\ 0 & \rho(-k) 
   \end{array}\right)
\end{equation}
The defining relations (\ref{rt-8})-(\ref{rt-10}), expressed for 
$r(k)$ and $t(k)$ with $S(k_{1},k_{2})$, then become:
\begin{eqnarray*}
 \!s_{12}(k_{1}-k_{2})\tau_{1}(k_1)\, s_{21}(k_{2}-k_{1}) 
 \tau_{2}(k_2)\! 
&=&\! \tau_{2}(k_2)s_{12}(k_{1}-k_{2}) \tau_{1}(k_1) s_{21}(k_{2}-k_{1})\ \
\\
\!s_{12}(k_{1}-k_{2}) \tau_{1}(k_1)s_{21}(k_{2}-k_{1})\rho_{2}(k_2)\! 
&=&\! \rho_{2}(k_2) s_{12}(k_{1}+k_{2}) \tau_{1}(k_1)
s_{21}(-k_{2}-k_{1})
\\
\!s_{12}(k_{1}-k_{2})\rho_{1}(k_1) s_{21}(k_{2}+k_{1}) \rho_{2}(k_2)\! 
&=&\! \rho_{2}(k_2) s_{12}(k_{1}+k_{2})\rho_{1}(k_1) s_{21}(k_{1}-k_{2})
\end{eqnarray*}
while the unitary relations (\ref{rt-12}) rewrite
\begin{eqnarray}
\tau(k)\tau(-k)+\rho(k)\rho(-k) &=& \II \,,\qquad \tau(k)^\dag=\tau(-k)\\
   \qquad \tau(k)\rho(-k)+\rho(k)\tau(-k) &=& 0 \,,\qquad \rho(k)^\dag=\rho(-k)
\end{eqnarray}
Note that this presentation has symmetrized the role of $\tau(k)$ and $\rho(k)$.

{}Finally, let us remark that this construction can obviously be 
applied to any $S$-matrix, translation invariant or not.

\end{document}